\documentclass[12pt]{article}

\usepackage{epsf}
\usepackage{epsfig}
\usepackage{cite}
\usepackage{amsmath}
\usepackage{graphics}

\begin{document}

\setlength{\textheight}{21.5cm}
\setlength{\oddsidemargin}{0.cm}
\setlength{\evensidemargin}{0.cm}
\setlength{\topmargin}{0.cm}
\setlength{\footskip}{1cm}
\setlength{\arraycolsep}{2pt}

\renewcommand{\thefootnote}{\#\arabic{footnote}}
\setcounter{footnote}{0}

\newcommand{\gtrsim}{ \mathop{}_{\textstyle \sim}^{\textstyle >} }
\newcommand{\lesssim}{ \mathop{}_{\textstyle \sim}^{\textstyle <} }
\newcommand{\rem}[1]{{\bf #1}}
\renewcommand{\thefootnote}{\fnsymbol{footnote}}
\setcounter{footnote}{0}
\def\thefootnote{\fnsymbol{footnote}}

\hfill {\tt April 2017}\\
\vskip .5in

\begin{center}

\bigskip
\bigskip

{\Large \bf Two Cosmic Coincidences for Minimal Standard Model with General Relativity}

\vskip .45in

{\bf Paul H. Frampton$^{(a)}$\footnote{paul.h.frampton@gmail.com}
 and Holger B. Nielsen$^{(b)}$\footnote{hbech@nbi.dk}}
 
 \vskip .3in
 
{\it $^{(a)}$ Department of Mathematics and Physics ``Ennio de Giorgi", \\
University of Salento, Lecce, Italy.}

\bigskip

{\it $^{(b)}$ The Niels Bohr Institute, University of Copenhagen,\\
Blegdamsvej 17, 2100 Copenhagen, Denmark.}

\end{center}

\vskip .4in
\begin{abstract}
It is said that there are no accidents or coincidences in physics. Within the minimal
standard model combined with general relativity we point out that there are three
exceptionally-long lifetimes which are consistent with being equal, and hence that there are
two unexplained coincidences. 
\end{abstract}

\renewcommand{\thepage}{\arabic{page}}
\setcounter{page}{1}
\renewcommand{\thefootnote}{\#\arabic{footnote}}

\newpage

\section{Introduction}

\noindent
The time scales or lifetimes associated with particle phenomenology
vary between those for strong interactions which may be as short as
$\sim 10^{-24}$ seconds up to exceedingly slow radioactive double
beta decay by weak interactions which has a measured lifetime up 
to $\sim 10^{21}$ years and some even slower undetected
processes such as proton decay.

\bigskip

\noindent
In classical general relativity no particular time scales stand out beyond
the present age of the universe $\sim 10^{10}$ yr which itself has no truly
fundamental significance. When we consider black holes of very high
mass, however, as we shall discuss there do emerge lifetimes
of an exceptionally long type.

\bigskip

\noindent
Buried in the gauge field theories which successfully describe 
particle phenomenology there are curious non-perturbative effects,
instantons, which give rise also to exceptional lifetimes
discussed in this article.

\bigskip

\noindent
A third type of unusual lifetime arises from the recently understood
origin of particle masses via the Englert-Brout-Higgs mechanism
because the measured values of the symmetry-breaking parameters
reveal that the physical vacuum is metastable with just such a long
lifetime.

\bigskip

\noindent
All three of these unusual lifetimes are consistent with being equal
or approximately so. This represents two coincidences between three
lifetimes which with present knowledge appear unrelated. 

\bigskip

\section{Lifetimes}

\subsection{Primordial Black Holes}

\noindent
It has been speculated \cite{PHF,CF} that the dark matter
in the Milky Way halo, as well as elsewhere in the universe,
is comprised of Primordial Black Holes (PBHs). In the local
halo the preferred mass range is between 25 and 625
solar masses corresponding to microlensing events
of stars in the Large Magellanic Cloud (LMC) and Small
Magellanic Cloud (SMC) with durations between 1 and
5 years.

\bigskip

\noindent
Elsewhere in the universe there exist far more massive black
holes, for example those at the cores of galaxies with masses
up to at least several times
$10^{10}M_{\odot}$ and we expect that these are
likewise primordial.

\bigskip

\noindent
The mathematics of hybrid inflation \cite{PBHformation,
PBHformation2}
shows the possiblity to
make PBHs all the way up to $10^{17}M_{\odot}$
if we allow the formation to wait until the end of the
radiation era at cosmic time $t \sim 50ky$. 
We note that PBHs with mass $10^5M_{\odot}$
are made at $t\sim 1s$ and thereafter the mass grows
linear in time. It seems
unlikely that such a PBH, with a mass far greater
than the largest cluster - indeed one millionth of the
total mass of the universe - really exists in our universe
but it is conceivable on general grounds that $10^{14}M_{\odot}$
PBHs might exist.

\bigskip

\noindent
Such a gigantic PBH might be unassociated with any cluster
or supercluster, and instead might simply drift alone. It is a
very interesting question whether there are limits on their
existence and how such limits might be established.

\bigskip

\noindent
The lifetime for decay of such a massive black hole is given
by a well-known formula, {\it e.g.} \cite{Carr},

\begin{equation}
\tau_{PBH} \simeq  10^{64} \left( \frac{M_{PBH}}{M_{\odot}} \right)^3  ~~ {\rm yr}.
\label{tauPBH}
\end{equation}

\noindent
Taking the largest mass PBH we have mentioned, $10^{14}M_{\odot}$, this means that the decay
lifetime satisfies $\tau_{PBH} \lesssim 10^{115}$ yr. We note that this lifetime
is not only much longer than the age $a_U$ of the universe, $a_U \sim 1.38 \times 10^{10}$ yr
but also much longer than the maximum estimate \cite{Sakharov}
for the proton decay lifetime $\tau_p$ which is $\tau_p \lesssim 10^{50}$ yr.

\bigskip

\noindent
The most massive known SMBH at a galactic core is in NGC4889,
336 Mly away, with mass
$2.1 \times 10^{10} M_{\odot}$. A heavier one, the most massive known black hole,
is in quasar S5 0014+81, 12.1Gly away, with mass $4 \times 10^{10} M_{\odot} \sim 10^{10.6} M_{\odot}$.
We have adopted $10^{14} M_{\odot}$ as a speculation for the most massive
PBH in the visible universe, only as about the geometric mean of the most
massive possible PBH and the most massive known BH. This precise value is not
important to our discussion, as from Eq.(\ref{tauPBH}) we see that any PBH
above $10^{12} M_{\odot}$ has an evaporation lifetime over one googol years.

\bigskip

\noindent
Such an extraordinarily long lifetime for the PBH decay might, at first, seem
unique within the context of the minimal standard model and general relativity.
But our purpose in the note is to emphasize that there is not only one but
actually two \cite{FH} other lifetimes in excess of $10^{100}$ yr which are not dissimilar from,
and consistent with, $\tau_{PBH}$.

\bigskip

\subsection{Instantons}

\noindent
Although perturbatively the standard model conserves baryon number and hence the
proton is stable, this ceases to be true when instantons are taken into account. At
zero temperature, the instanton effects are exponentially suppressed and
an estimate of the proton lifetime $\tau_{p,inst}$ for instanton-induced decay\cite{Hooft,Hooft2} is
much much longer than the value of $\tau_p$ cite above, and is given by

\begin{equation}
\tau_{p,inst} \simeq {\rm exp} \left( + 4 \pi {\rm sin}^2 \Theta_W / \alpha_{em} \right)
\label{taupinstanton}
\end{equation}

\noindent
This estimate, Eq.(\ref{taupinstanton}), has a numerically-irrelevant prefactor and gives an instanton-induced proton decay lifetime
with a lower limit of at least $\tau_{p,inst}  > 10^{100}$ y.

\bigskip

\noindent
This is not a prediction which can be directly tested since it is empirically indistinguishable
from absolute stability, and therefore seems, {\it prima facie}, academic. Nevertheless, 
this lifetime $\tau_{p,inst}$ is consistent with $\tau_{PBH}$.

\bigskip

\subsection{Vacuum Metastability}

\noindent
In 2012 the Higgs boson was discovered with a mass of $M_H\simeq126$ GeV and using the
best measured value for the top quark mass $M_t$ an analysis \cite{BDGGSSS} of the
effective potential of the minimal standard model with only one complex scalar doublet
revealed that the physical vacuum in which we live is metastable with an exceptionally
long lifetime.

\bigskip

\noindent
This metastability is sensitively dependent on $M_H$ and $M_t$ and only a small variation
from the meausred values of these two quantities would provide absolute stability. As
examples if we fix $M_H$ at its physical value and reduce the top mass to $171$ GeV
from the correct value $\sim 173.36$ GeV, or if we fix $M_t$ at its correct value
and increase $M_H$ to  $130$ GeV from its physical value $\sim 125.66$ GeV,
the vacuum would be stable.

\bigskip

\noindent
The interpretation of this phenomenon remains unclear, and with the physical $M_H$
and $M_t$ additional states
can render the vacuum stable as illustrated in \cite{Laperashvili}. 

\bigskip

\noindent
If we retain only the minimal number of states in the standard model
the vacuum lifetime is a similarly exceptional time of $\tau_{vac} \sim 10^{100}$ yr.

\bigskip

\section{Discussion}

\noindent
Two of the three lifetimes we have mentioned were put equal in
\cite{FH}, $\tau_{p,inst} = \tau_{vac}$ with, as only variable,
the Higgs mass $M_H$. This led to the result $M_H = 126\pm 4$ GeV
and could be regarded as a prediction for the otherwise unexplained 
Higgs mass. Putting the two lifetimes equal does not yet have another 
justification.

\bigskip

\noindent
A similar argument could be used for $\tau_{PBH}$ for the supermassive
PBHs in the universe to argue about the value of the highest mass black
hole in the universe, although again the assumption that the maximum
$\tau_{PBH}$ is equal to one of the other two lifetimes 
$\tau_{vac}$ or $\tau_{p,inst}$ is necessary.

\bigskip

\noindent
It is our belief that the two ``cosmic" coincidences between the three
exceptional lifetimes might have a deep significance for theoretical physics
and may play a role in a more complete theory.

\bigskip

\begin{center}
\section*{Acknowledgement}

\end{center}

\noindent This work was initiated at the MIAMI2016 Conference held in Fort Lauderdale
in December 2016 and organized by T.L. Curtright of the University of Miami.

\end{document}